\pdfoutput=1
\documentclass[conference]{IEEEtran}

\usepackage{url}
\usepackage{graphicx}

\ifCLASSINFOpdf
\else
\fi

\begin{document}

\newcommand\copyrighttext{
	\Huge {IEEE Copyright Notice} \\ \\
	\large {Copyright (c) 2017 IEEE \\
		Personal use of this material is permitted. Permission from IEEE must be obtained for all other uses, in any current or future media, including reprinting/republishing this material for advertising or promotional purposes, creating new collective works, for resale or redistribution to servers or lists, or reuse of any copyrighted component of this work in other works.} \\ \\
	
	{\Large Published in: 2017 IEEE International Conference on AI \& Mobile Services
		(IEEE AIMS 2017), June 25-30, 2017} \\ \\ 
	DOI: 10.1109/AIMS.2017.22 \\ \\
	\begin{small}
		Preprint from: \url{https://arxiv.org/abs/1705.10512}\\
		Print at: \url{https://doi.org/10.1109/AIMS.2017.22}
	\end{small}
		
	\vspace{2cm}
		
	Cite as:\\
	\includegraphics[trim={2.075cm 23cm 9cm 3.1cm},clip]{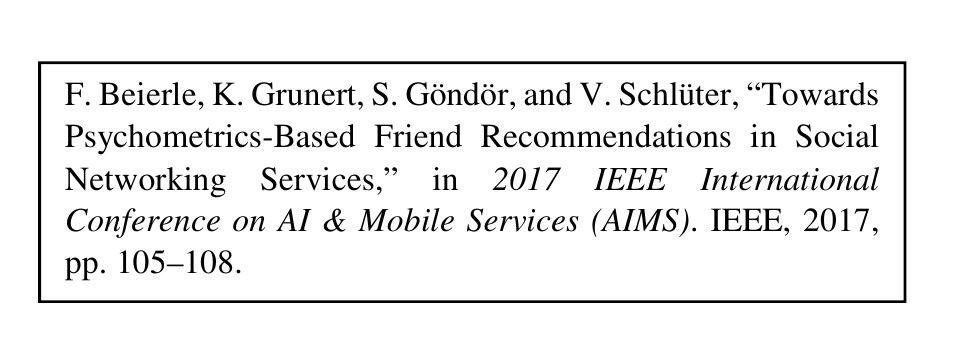}

	\vspace{1.5cm}
	
	BibTeX:\\ \\
	\includegraphics[trim={2.25cm 22cm 2cm 4cm},clip]{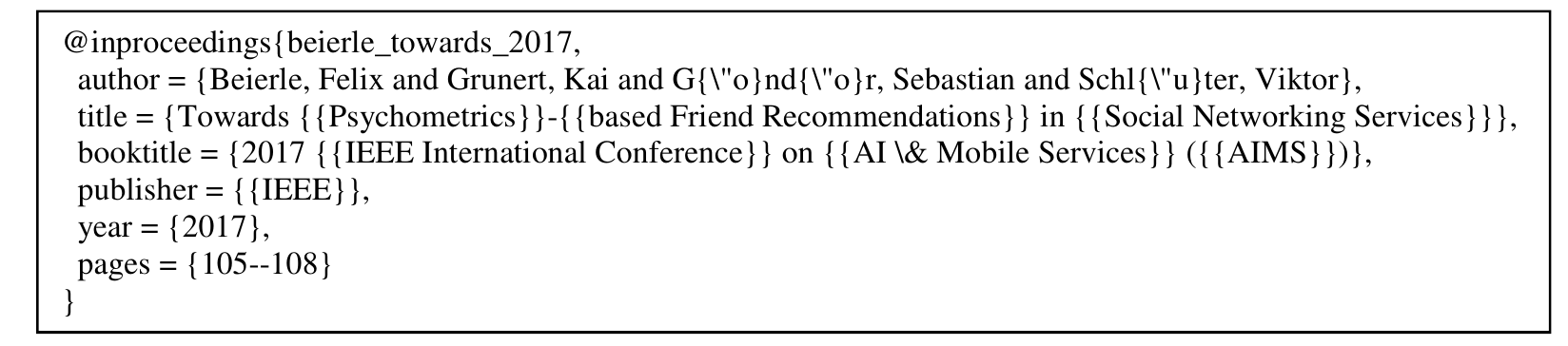}
}

\twocolumn[
\begin{@twocolumnfalse}
	\copyrighttext
\end{@twocolumnfalse}
]

\title{Towards Psychometrics-based Friend Recommendations in Social Networking Services}

\author{
	\IEEEauthorblockN{
		Felix Beierle\IEEEauthorrefmark{1},
		Kai Grunert\IEEEauthorrefmark{1},
		Sebastian G\"ond\"or\IEEEauthorrefmark{1},
		Viktor Schl\"uter\IEEEauthorrefmark{2}}
	\IEEEauthorblockA{
		\IEEEauthorrefmark{1}
			Service-centric Networking\\
			Telekom Innovation Laboratories / Technische Universit\"at Berlin\\
			Berlin, Germany\\
			\{beierle, kai.grunert, sebastian.goendoer\}@tu-berlin.de}
	\IEEEauthorblockA{
		\IEEEauthorrefmark{2}
			Technische Universit\"at Berlin, Berlin, Germany\\
			viktor.a.schlueter@campus.tu-berlin.de}
}

\maketitle

\begin{abstract}

Two of the defining elements of Social Networking Services are the
social profile, containing information about the user, and the social graph,
containing information about the connections between users.
Social Networking Services are used to connect to known people
as well as to discover new contacts.
Current friend recommendation mechanisms typically utilize the social graph.
In this paper, we argue that
psychometrics, the field of measuring personality traits,
can help make meaningful friend recommendations based on
an extended social profile containing collected smartphone sensor data.
This will support the development of
highly distributed Social Networking Services without central knowledge of
the social graph.

\end{abstract}

\IEEEpeerreviewmaketitle

\section{Introduction}

Social Networking Services (SNSs) are one of the most used services
on the World Wide Web \cite{greenwood_social_2016}.
Two typical elements of a SNS are the \emph{social profile}, containing information
about a user, for example her interests, and the \emph{social graph},
containing information about the connections between users.
In our previous {\tiny }work, we argued that the smartphone is the optimal social
networking device \cite{beierle_towards_2015}.
It typically has only one user and, 
with recent developments in smartphone sensor technologies and available APIs,
more and more personal data -- like location traces, most frequently used apps,
etc.\ -- is available that could potentially extend existing social profiles.

One of the typical applications in SNSs are friend recommendations.
When recommending new connections in an SNS, typically, the
social graph is utilized \cite{yin_unified_2010}.
While
doing so
enables the
incorporation of graph-based
properties like the number of mutual friends,
there are also studies that
look into the similarity of attributes of neighboring nodes,
thus incorporating the social profile in the recommendation process
\cite{mohajireen_relational_2011}.
The basis for the cited studies about friend recommendations
is the insight that \emph{homophily}
-- the tendency for people to associate themselves with people
who are similar to them --
is structuring any type of network \cite{mcpherson_birds_2001}.
Looking further into the fields of psychology and social sciences,
\emph{psychometrics}, the academic field that deals with
measuring psychological personality traits,
seems like a promising research area providing results that could help
improve friend recommendations in SNSs.
Recently, the company \emph{Cambridge Analytica}
was in the media because of their alleged success
in utilizing psychometrics in targeted political campaign advertisements \cite{blakely_data_2016},
though their impact on the campaign remains somewhat unclear \cite{confessore_data_2017}.
Although the use case is different
-- targeted advertising instead of friend recommendation --
this shows the potential of applying psychology research results to other fields.
In this paper, we argue that combining
current smartphone technologies with findings from psychometrics
will enable meaningful friend recommendations based on
social profiles without requiring knowledge of the social graph.
Our main contributions of this short paper is
a thorough analysis of the theoretical background of psychometrics in relation
to SNSs and mobile devices, including a proposal of
how to integrate the insights into friend recommendations
in SNSs.

\section{Analysis and Concept}

In this section, we give a detailed analysis of relevant
work related to psychometrics, social networking, and smartphone usage.
In Section~\ref{sec:2:psych},
we give a literature review on how and why people actually connect
with each other in (offline) social networks.
We outline the concepts of homophily and personality
from psychology and social sciences.
In order to ensure that the same concepts hold true in
SNSs, we look into existing research on SNSs and personality
in Section~\ref{sec:2:socialnetworkpersonality}.
In Section~\ref{sec:2:smartphonepersonality},
we investigate the relationship between smartphone usage and personality.
As we will deal with attributes of users rather then
existing connections between users, we will look at existing
definitions and components of social profiles in
Section~\ref{sec:2:definition}.

\subsection{Psychology and Social Sciences}
\label{sec:2:psych}

In this section, we want to investigate two concepts from psychology
and social sciences: homophily and personality.
Those two concepts will help us conceptualize
the parameters we need in a system for
psychometrics-based friend recommendations.
Furthermore, it will
answer the questions
"When do people become friends?", i.e., "When do people create edges in a social graph?",
which are necessary to be asked in social networking.

Homophily is the concept that people tend to associate themselves with
other people that are similar to them.
According to McPherson et al., this principle structures network
ties of every type, including friendship, work, or partnership \cite{mcpherson_birds_2001}.
Some of the categories in which people have homophilic contacts are
ethnicity, age, education, and gender.

The social profile is representing a user.
The personality of a person influences
a multitude of aspects, e.g., job performance, satisfaction, or romantic success
\cite{golbeck_predicting_2011}
and is a "key determinant for the friendship formation process"
\cite{burgess_school_2011}.
One of the established ways to talk and research about personality is the so-called
Big Five or Five-Factor model \cite{tupes_recurrent_1992,mccrae_introduction_1992}.
The five personality factors spell the acronym OCEAN and are
\emph{openness to experience},
\emph{conscientiousness},
\emph{extraversion},
\emph{agreeableness}, and
\emph{neuroticism}.
In their study, Selfhout et al.\
show the importance of homophily for friendship networks \cite{selfhout_emerging_2010}.
For three of the five factors (openness to experience, extraversion, and agreeableness),
they conclude that people tend to select friends with similar levels of those traits.

Additionally to friendship, there are several studies finding correlations between
different aspects of everyday life and the five factors.
Especially interesting for social networking related questions
is the correlations between the five factors and preferences or interests.
\cite{rawlings_music_1997} and
\cite{rentfrow_re_2003}
are two of the studies that find correlations between
personality and the music the persons prefer to listen to.
As we will show in Section~\ref{sec:2:definition},
music preference
is a typical element for a social profile.
It is the most commonly filled attribute in publicly
accessible Facebook profiles after gender \cite{farahbakhsh_analysis_2013}.

\subsection{Social Networking Services and Personality}
\label{sec:2:socialnetworkpersonality}

Several studies suggest that
the findings about (offline) social networks
are also valid when dealing with SNSs.
Liu claims the social profile is a "performance" by the user
who expresses herself by crafting the profile \cite{liu_social_2007}.
While this might be true, various studies show that this does not
imply that this "performance" distorts the personality that is expressed
in the profile.
For example, Back et al.\ conclude in their study that "Facebook Profiles
Reflect Actual Personality, Not Self-Idealization," as the title of their paper
indicates \cite{back_facebook_2010}.
In their study, Goldbeck et al.\ show that 
Facebook profiles can be used to predict personality \cite{golbeck_predicting_2011}.
Another study comes to the same conclusion and shows
"that Facebook-based personality impressions
show some consensus for all Big Five dimensions"~\cite{gosling_personality_2007}.

\subsection {Smartphone Usage and Personality}
\label{sec:2:smartphonepersonality}

Some studies on pre-smartphone-era cell phones found
correlations between personality traits and mobile phone usage.
For example, these are the results of a study done on the general
use of mobile phones (calls, text messages, changing ringtones and wallpapers)
\cite{butt_personality_2008},
as well as of a study about using
mobile phone games~\cite{phillips_personality_2006}.
While these studies were based on self reports by users,
Chittaranjan et al.\ conducted two user studies in which they collected usage
data on Nokia N95 phones \cite{chittaranjan_whos_2011,chittaranjan_mining_2013}.
In those studies, the authors were looking at
Bluetooth scan data, call logs, text messages,
calling profiles, and application usage.
At the time the study was conducted, apps were not as common as nowadays with Android and iOS.
The authors state that "features derived from the App Logs were sparse due to
the low frequency of usage of some of the applications" \cite{chittaranjan_mining_2013}.
It will be interesting to compare the results from their study to
a new study where the usage of apps is commonplace.
The results of the cited studies indicate that
"several aggregated features obtained from smartphone usage
data can be indicators of the Big-Five traits" \cite{chittaranjan_mining_2013}.

In a more recent study, Lane and Manner showed relations
between the usage of apps and the five personality dimensions
\cite{lane_influence_2012}.
Apps were categorized in different application types:
communication, games, multimedia, productivity, travel, and utilities.
Overall, the referenced studies indicate strong
correlations between smartphone usage behavior and personality traits.

\subsection{Social Profiles}
\label{sec:2:definition}

The social profile is one of the central elements of SNSs.
In Boyd and Ellison's definition of \emph{Social Network Sites}, 
the "public or semi-public profile within a bounded system" is the
first defining element, and the "backbone" of the SNS \cite{boyd_social_2007}.
Typical elements of a social profile are
"age, location, interests, and an 'about me' section," and a photo.
In \cite{richter_functions_2008}, the social profile is the first defining functionality
of an SNS. Here, the authors call the functionality "identity management,"
as the profile is a "representation of the own person."
In a survey paper about SNSs,
the social profile is described as the "core" of an Online
Social Network \cite{heidemann_online_2012}.
Another recent survey describes the
creation and maintenance of user profiles as the
"basic functionality" of SNSs \cite{paul_survey_2014}.
In \cite{gondor_sonic:_2015}, several SNSs from different categories,
like general (e.g. Facebook), business oriented (e.g. LinkedIn)
or special purpose (e.g. Twitter), were analyzed.
The social profile is an element that is present in all of those SNSs.
Rohani and Hock state that the type of information included in social profiles
differs between different SNSs \cite{rohani_social_2009}.
In their analysis of publicly disclosed Facebook profile information,
Farahbakhsh et al.\ distinguish between personal and
interest-based attributes \cite{farahbakhsh_analysis_2013}.
Personal attributes include a friend list,
current city, hometown, gender, birthday, employers, college, and high school.
Interest-based attributes are music, movie, book, television, games, teams, sports,
athletes, activities, interests, and inspirations.
Lampe et al.\ distinguish between three different types of
information:
referents, interests, and contact \cite{lampe_familiar_2007}.
Referents include verifiable attributes:
hometown, high school, residence, concentration.
Contact information are also verifiable, for example
website, email, address, or birthday.
As the authors indicate, interests are less verifiable.
Interests include an 'about me' section,
favorite music, movies, TV shows, books, quotes, and political views.
As the cited studies in Section~\ref{sec:2:smartphonepersonality} 
and also social networking related studies (e.g., \cite{bao_recommendations_2015})
suggest, more detailed user data additional to the data typically available
in a social profile can help improve recommendations in SNSs.

\section{Concept and Prototype}
\label{sec:2:concept}

Research in psychology and social sciences indicates that homophily
in age, education, etc., as well as in personality traits, is a strong
indicator for friendship, i.e., for the creation of an edge in a social graph.
Several of the aforementioned studies conclude with findings about correlations
between smartphone usage and personality traits.
Combining those insights, in order to
make meaningful recommendations for new connections in an SNS,
we can recommend users that show a similar behavioral pattern
with their smarthphones.
As the existing studies suggest, this will indicate the similarity
of their personality traits.
Doing this, we do not necessarily need to know which
behavior indicates which personality trait.

Such a mechanism for friend recommendations can have several benefits:
(1) By logging information about the smartphone usage behavior
(a lot of which can be done unobtrusively, see \cite{beierle_privacy-aware_2016}),
we could automatically set up or update an existing social profile in an SNS,
eliminating or reducing the tedious task to keep such a profile up to date
\cite{lampe_familiar_2007}.
(2) When fully relying on social profile data for friend recommendations,
the social graph is not needed in the recommendation process.
This will enable the development of highly
distributed SNSs, for example in device-to-device scenarios
where two randomly meeting people could determine
-- without contacting some centralized server --
how similar they are and thus are recommended to be friends.
(3) By collecting highly personalized user data, further studies
in the fields of psychometrics could be possible.

We are implementing an Android prototype.
With the \emph{Google Awareness API}, Google offers
to get the user's current
time, location, nearby places, nearby beacons, headphone state, activity,
and weather.\footnote{\url{https://developers.google.com/awareness/}}
Android also enables developers to
retrieve a list of the most frequently used apps.
Most music players broadcast what the user is listening to,
so other apps can retrieve this information.
The available APIs and mechanisms allow for an unobtrusive collection
of user data on Android smartphones.
After collecting the mentioned data,
in order to estimate their similarity,
two users can share their data in a device-to-device manner
by utilizing appropriate data structures
and technologies like Bluetooth, Wi-Fi Direct, or Wi-Fi Aware.

\section{Related Work}

In this section,
we review related work on smartphone sensor data collection
(Section \ref{sec:relwork:smartphonedata})
and on link prediction in SNSs
(Section \ref{sec:relwork:linkprediction}).

\subsection{Smartphone Data Collection}
\label{sec:relwork:smartphonedata}

In \cite{pejovic_anticipatory_2015}, the authors survey
the state of the art of "Anticipatory Mobile Computing,"
describing how the advances in mobile technology will
enable predicting future contexts and acting on it.
This field has some similarities to our work,
especially with respect to collecting and using sensor data,
but it does not focus on social networking.

In \cite{hashemi_user_2016}, the authors present a framework called "BaranC"
for monitoring and analyzing digital interaction of users with their smartphones.
The architecture uses cloud technologies to analyze data and thus
raises privacy concerns.
The goal of BaranC is not social networking but offering personalized services.
In another work by the same authors, an application utilizing their framework
is presented \cite{hashemi_next_2016},
predicting the next application a user will use.

Wang et al.\ present a system that collects a multitude of sensor data from
mobile devices and queries the users with questionnaires.
The collected data is then used to accurately predict the GPA of the
undergraduate students who participated \cite{wang_smartgpa:_2015}.
Xiong et al.\ present a system for social sciences studies
that collects sensor data and enables researchers to create surveys for study
participants \cite{xiong_sensus:_2016}.
Again, in both those cases, the developed concepts did not
focus on social networking.

\subsection{Link Prediction}
\label{sec:relwork:linkprediction}

One of the common ways to research about links in social networks is \emph{link prediction}.
The key difference to our work is that here, the social graph is used to
calculate the prediction or recommendation,
while our approach is also feasible in device-to-device mobile SNS scenarios
where the social graph is not known.
Yin et al.\ analyzed links in social networks based on "intuition-based" aspects:
homophily (shared attributes), rarity (matching uncommon attributes), social influence (more likely to link to person that shares attributes with existing friends), common friendship (mutual friends), social closeness (being close to each other in the social graph), preferential attachment (more likely to link to a popular person) \cite{yin_unified_2010}.
Most aspects focus on the social graph or global knowledge about
attribute distribution (in the case of rarity).
In the work by Mohajureen et al., the authors
use the attributes of neighboring nodes in the
friend recommendation process \cite{mohajireen_relational_2011}.
For this algorithm to work, the social graph as well as the features of
each user have to be available.

A somewhat special case of link prediction or friend recommendation is
described in \emph{WhozThat?} \cite{beach_whozthat?_2008}.
Here, the idea is to retrieve information about another person you just met.
Via Bluetooth, user handles from an SNS are exchanged and data about the other person
can be retrieved from that SNS.
As described in Section~\ref{sec:2:concept},
by following our concept,
the same scenario can be realized
in a distributed manner
without contacting existing centralized SNS.

\section{Conclusion and Future Work}

In this paper, we proposed the extension of
social profiles with smartphone sensor data.
We showed that research results from the field of psychometrics
suggest that we then can calculate relevant friend recommendations
based on those profiles, without utilizing the social graph.
This will enable recommendations in highly distributed SNSs.

Future work includes conducting a user study with our prototype
to confirm our conclusions.
Furthermore, in the device-to-device SNS scenario, the
issue of privacy-awareness will be further investigated
so that similarity estimations between users are possible
without sharing all the collected sensitive sensor data.

\section*{Acknowledgment}
This work has received funding from the European Union's Horizon 2020
research and innovation program under grant agreement No 645342, project reTHINK
and from project DYNAMIC\footnote{\url{http://www.dynamic-project.de}} (grant No 01IS12056), which is funded
as part of the Software Campus initiative by the German
Federal Ministry of Education and Research (BMBF).

\bibliographystyle{IEEEtran}

\end{document}